\documentclass[aps,prb,preprint,groupedaddress]{revtex4-1}
\usepackage{graphicx,amsmath}
\usepackage{epstopdf}

\begin{document}


\title{Magnetic structure of hexagonal YMnO$_3$ and LuMnO$_3$
from a microscopic point of view}


\author{I. V. Solovyev}
\email{SOLOVYEV.Igor@nims.go.jp}
\affiliation{Computational Materials Science Unit,
National Institute for Materials Science, 1-2-1 Sengen, Tsukuba,
Ibaraki 305-0047, Japan}
\author{M. V. Valentyuk}
\altaffiliation[Temporarily at ]
{Institute of Theoretical Physics, University of Hamburg, Jungiusstrasse 9, 20355 Hamburg, Germany}
\author{V. V. Mazurenko}
\affiliation{
Department of Theoretical Physics and Applied Mathematics, Ural Federal University,
Mira str. 19, 620002 Ekaterinburg, Russia}


\date{\today}

\begin{abstract}
The aim of this work is to unravel a basic microscopic picture
behind complex magnetic properties of hexagonal manganites.
For these purposes, we consider two characteristic compounds: YMnO$_3$ and LuMnO$_3$,
which form different magnetic structures in the ground state
($P\underline{6}_3c\underline{m}$ and $P\underline{6}_3\underline{c}m$, respectively).
First, we establish an electronic low-energy model, which describes the behavior of the Mn $3d$ bands of YMnO$_3$ and LuMnO$_3$,
and derive parameters of this model from the first-principles calculations.
From the solution of this model, we conclude that, despite strong frustration effects in the hexagonal lattice,
the relativistic spin-orbit interactions lift the degeneracy of the magnetic ground state so that
the experimentally observed magnetic structures are successfully reproduced by the low-energy model.
Then, we analyze this result in terms of interatomic magnetic interactions, which were computed using
different approximations (starting from the model Hamiltonian as well as directly from the first-principles electronic
structure calculations in the local-spin-density approximation). We argue that
the main reason why
YMnO$_3$ and LuMnO$_3$ tend to form different magnetic structures
is related to the behavior of the single-ion anisotropy,
which reflects the directional dependence of the lattice distortion: namely, the expansion and contraction of the Mn-trimers,
which take place in YMnO$_3$ and LuMnO$_3$, respectively. On the other hand, the magnetic coupling between the planes
is controlled by the next-nearest-neighbor interactions, which are less sensitive to the
direction of the trimerization.
In the $P\underline{6}_3c\underline{m}$ structure of YMnO$_3$,
the Dzyaloshinskii-Moriya interactions lead to the spin canting out of the hexagonal plane -- in the same
direction as the single-ion anisotropy.
Finally, using the Berry-phase formalism,
we evaluate the magnetic-state dependence of the ferroelectric polarization, and discuss potential applications
of the latter in magnetoelectric switching phenomena.
\end{abstract}


\maketitle

\section{\label{Intro} Introduction}

  Hexagonal manganites (the space group $P6_3cm$)
are one of canonical examples of multiferroics, which have attracted
an enormous attention recently.
The coexistence of ferroelectricity and magnetism in such systems provides a unique possibility
for manipulating the charges by applying a magnetic field and the spins by applying a voltage,
which is crucially important for the construction of new forms of multifunctional devices.\cite{Mostovoy}
To this end,
the direct magnetic phase control by static electric field was realized in HoMnO$_3$.\cite{Fiebig2004}
The interplay between the ferroelectric activity and the magnetic order was also
demonstrated in YMnO$_3$ and LuMnO$_3$ with the measurements of the dielectric constant and the loss tangent,
which were shown to exhibit clear anomalies around the N\'eel temperature ($T_{\rm N}$$=$ 75 K and 88 K in
YMnO$_3$ and LuMnO$_3$, respectively),\cite{Huang,Katsufuji}
even despite the fact that the ferroelectric transition itself occurred at much higher temperature
($T_{\rm C}$$\sim 880$ K).\cite{ChoiNM}
Another spectacular example is the coupling of magnetic and ferroelectric domains, which was
visualized in YMnO$_3$ by using optical second harmonic generation technique.\cite{Fiebig2002}
Furthermore,
the magnetic transition in YMnO$_3$ and LuMnO$_3$ is accompanied by a distinct change of the atomic positions.\cite{LeeNature}
Thus, the experimental data clearly demonstrates the existence of a strong coupling amongst electric, magnetic, and
lattice degrees of freedom in these hexagonal manganites.

  The magnetic frustration is one of the key concepts of multiferroic materials, which may assist
the inversion symmetry breaking and, in a number of cases, be even responsible for such a breaking.\cite{PRB11}
In this respect, the hexagonal lattice is not an exception, and
is typically regarded as a playground for studying
the magnetic frustration effects. However, it is also the main complication, hampering the
theoretical understanding of multiferroic effects in hexagonal compounds, even
despite the fact that the high-spin state
($S$$=$$2$), realized in manganites, is typically regarded as an ``easy case'' for such theoretical analysis,
where the classical the spin fluctuations dominate over the quantum ones.
Nevertheless,
the ground state of classical spins in the hexagonal lattice is expected to be highly degenerate.
Different signs of spin fluctuations, apparently originating from this degeneracy,
were indeed observed in the neutron scattering experiments, even below $T_{\rm N}$.\cite{Perring,Sato}
Another evidence of spin fluctuations, which is also related to the quasi-two-dimensional character of
magnetic interactions,
is the large ratio of the Curie-Weiss temperature ($\theta_{\rm CW}$)
to $T_{\rm N}$ (about 7 in YMnO$_3$).\cite{Perring}

  The degeneracy can be lifted by lattice distortions and, in this context, plenty of attention
is paid to the so-called trimerization instability, inherent to the $P6_3cm$ structure.\cite{LeeNature,Park}
However, the trimerization alone does not lift the frustration of isotropic exchange interactions.
In this sense, the situation is fundamentally different from
the exchange striction effect, which accompanies the formation of the $E$-type antiferromagnetic (AFM)
state in the orthorhombic YMnO$_3$ and which lifts the frustration of nearest-neighbor (NN) interactions.\cite{Okuyama}
Nevertheless,
the trimerization
can interplay with the relativistic spin-orbit (SO) coupling
and, in this way,
give rise to new anisotropic interactions,
which can lift the degeneracy and stabilize some individual magnetic structure with the
well-defined symmetry.
Such structures were detected in the experiments on
the neutron diffraction (Ref.~\onlinecite{Park,Munoz,Brown}) and optical second harmonic generation (Ref.~\onlinecite{Fiebig2000}).
In a number of cases (e.g., in LuMnO$_3$), there can be several magnetic structures, coexisting in
a narrow temperature range.\cite{Fiebig2000}
In short, despite difficulties, there is an enormous experimental progress in
the identification of magnetic structures of hexagonal manganites,
resulting from a delicate balance between lattice distortion, SO interaction, and frustration effects.

  The microscopic understanding of rich magnetic properties of the hexagonal manganites is still rather limited.
To begin with, there is no clear microscopic model, which would
explain the origin of basic magnetic structures of hexagonal manganites,
and why different manganites tend to form different magnetic structures.
Basically,
it is only known how the trimerization affects the NN isotropic interactions.\cite{Park}
The presence of single-ion anisotropy and Dzyaloshinskii-Moriya (DM) interactions is, of course, anticipated. However,
it is absolutely not clear how all these effects come together to form a variety of magnetic structures,
realized in
the hexagonal manganites.

  In this paper, we will try to answer some of these questions. For these purposes, we
consider two characteristic manganites: YMnO$_3$ and LuMnO$_3$, which form different
magnetic structures in the ground state: $P\underline{6}_3c\underline{m}$ and $P\underline{6}_3\underline{c}m$, respectively
(in the International notations, where each underlined
symbol means that given symmetry operation is combined with the time inversion).
We will show that this difference can be naturally related to different directions of the trimerization:
expansion and contraction of the Mn-trimer, which takes place
in YMnO$_3$ and LuMnO$_3$,
respectively.
In our study, we start from the first-principles electronic structure calculations.
First, we construct a low-energy electronic model, which captures details of the magnetic structure
and correctly reproduces the magnetic ground state of YMnO$_3$ and LuMnO$_3$.
Then, we analyze these results by further transforming the electronic model into the spin one
and elucidating which magnetic interaction is responsible for each detail of the magnetic structure.
We will also consider the `temperature effect', associated with the temperature change of the
experimental crystal structure, and show that above $T_{\rm N}$ it gradually diminishes the
anisotropic interactions.

  The rest of the paper is organized as follows.
All methodological aspects, such as construction of the electronic model and
calculation of magnetic interactions, are discussed in Sec.~\ref{Method}.
Results of solution of the electronic model in the Hartree-Fock (HF) approximation are presented in Sec.~\ref{MStructures}.
In Sec.~\ref{MInteractions}, we give a detailed analysis of the obtained results in terms of
magnetic interactions, which were computed using different starting points.
In Sec.~\ref{Polarization}, we discuss the magnetic part of the ferroelectric polarization and propose
how it can be controlled by switching the magnetic state.
Finally, a brief summary of the work is given in Sec.~\ref{Conclusions}.

\section{\label{Method} Method}

  Since our goal is the construction of microscopic theory for the magnetic properties of
YMnO$_3$ and LuMnO$_3$, we first adopt the low-energy model, which would provide a realistic description for the Mn $3d$ bands
of these compounds:
\begin{equation}
\hat{\cal{H}}  =  \sum_{ij} \sum_{\alpha \beta}
t_{ij}^{\alpha \beta}\hat{c}^\dagger_{i\alpha}
\hat{c}^{\phantom{\dagger}}_{j\beta} +
  \frac{1}{2}
\sum_{i}  \sum_{ \{ \alpha \} }
U_{\alpha \beta \gamma \delta} \hat{c}^\dagger_{i\alpha} \hat{c}^\dagger_{i\gamma}
\hat{c}^{\phantom{\dagger}}_{i\beta}
\hat{c}^{\phantom{\dagger}}_{i\delta}.
\label{eqn.ManyBodyH}
\end{equation}
The model is constructed in the basis of Wannier orbitals, using the input from
the first-principles electronic structure calculations.
Each Wannier orbital is denoted by the
Greek symbol, which itself is the combination of spin
($s$$=$ $\uparrow$ or $\downarrow$) and orbital
($m$$=$ $xy$, $yz$, $3z^2$$-$$r^2$, $zx$, or $x^2$$-$$y^2$) variables.
Since the Mn $3d$ bands in hexagonal manganites are well separated from the rest of the spectrum,\cite{Park}
the construction of the model Hamiltonian (\ref{eqn.ManyBodyH}) is rather straightforward. The corresponding procedure
can be found in Ref.~\onlinecite{review2008}.

  All calculations have been performed using experimental parameters of the crystal structure, measured
at 10 K and 300 K (Ref.~\onlinecite{LeeNature}, Supplementary Information),
i.e. well below and above the magnetic transition point.
The experimental space group $P6_3cm$ has 12 symmetry operations, which can be generated by the
mirror reflection $x$$\rightarrow$$-$$x$, $m_x$, and the $60^\circ$-degree rotation
around the $z$-axis, combined with the half of the hexagonal translation,
$\{ C^6_z | {\bf c}/2 \}$.

  The crystal-field splitting, obtained from the diagonalization of the
site-diagonal part of $\hat{t}_{ij} = \| t_{ij}^{\alpha \beta} \|$ (without spin-orbit coupling),
is very similar in YMnO$_3$ and LuMnO$_3$. For example, if one uses the 10 K structure, we obtain the following
scheme of the atomic levels: $-$$0.54$, $-$$0.43$, $-$$0.29$, $-$$0.24$, and $1.50$ eV in the case of YMnO$_3$, and
$-$$0.60$, $-$$0.49$, $-$$0.25$, $-$$0.24$, and $1.57$ eV in the case of LuMnO$_3$. The use of the 300 K structure
yields similar results. Clearly, the crystal field tends to stabilize four atomic orbitals, which are
separated from the fifth one by the large energy gap. Such a scheme of the crystal-field splitting is
consistent with the formal $d^4$ configuration of the Mn-ions, which is subjected to the Jahn-Teller instability.
The fifth (unoccupied) orbital is of predominantly $3z^2$$-$$r^2$ symmetry. The off-diagonal elements of
$\hat{t}_{ij} = \| t_{ij}^{\alpha \beta} \|$ with respect to the site indices stand for the transfer integrals.
They are listed in Ref.~\onlinecite{SM}. The value of the
screened Coulomb interaction $U$ (defined as radial Slater's integral $F^0$) is about $2.6$ eV
for all considered systems. The intraatomic exchange (Hund's) coupling $J_{\rm H}$ is about $0.9$ eV, which is practically unscreened.
The full matrices of screened Coulomb interactions
can be also found in Ref.~\onlinecite{SM}.

  After the construction, the model is solved in the HF approximation.\cite{review2008}
This procedure appears to be extremely useful, especially for the search of the magnetic ground state.
Typically, in frustrated magnetic systems, we are dealing with the competition of several magnetic interactions
of the both relativistic and non-relativistic origin. Therefore, even HF calculations for the
relatively simple model (\ref{eqn.ManyBodyH}) can be very time consuming, because they may require tends of thousands
of iterations. In such a situation, the full scale electronic structure calculations are simply unaffordable.
Since the degeneracy of the ground state is lifted by the lattice distortion, the HF approximation
appears to be a good starting point for the analysis of the equilibrium magnetic properties.\cite{review2008}

  Of course, the model (\ref{eqn.ManyBodyH}) is not perfect, because it does not explicitly include the
oxygen band, which can be important for the quantitative analysis of magnetic properties of manganites.
Therefore, whenever possible, we check results of our model analysis by direct calculations in the
local-spin-density approximation (LSDA). For these purposes, we use the tight-binding
linear muffin-tin-orbital method (in the following we will refer to such calculations as `LMTO calculations').\cite{LMTO}
Hopefully, in both cases we can employ the same strategy
for calculations of magnetic interactions, which is based on the local force theorem and the
Green's function technique. Namely, the isotropic exchange interactions ($J_{ij}$) can be obtained in the
second order perturbation-theory expansion for the infinitesimal spin rotations,\cite{Liechtenstein}
antisymmetric DM interactions (${\bf d}_{ij}$) -- by considering mixed type
perturbation with respect to the rotations and the relativistic SO coupling,\cite{PRL96,DMI,MnCuN} and the
single-ion anisotropy tensors ($\hat{\tau}_i$) -- in the second order with respect to the SO interaction \cite{PRB95}.

\begin{figure}[h!]
\begin{center}
\includegraphics[width=8cm]{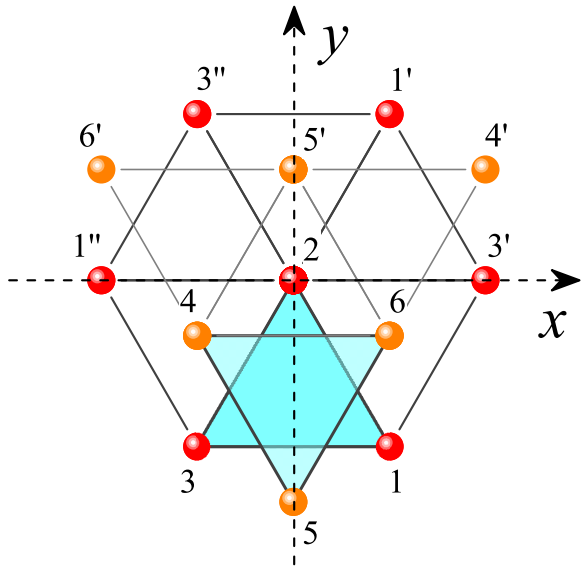}
\end{center}
\caption{\label{fig.AtomicPositions}
(Color online)
Relative positions of Mn-sites in the hexagonal $P6_3cm$ structure: the atoms
located in the plane $z$$=$$0$ are indicated by the red (dark) spheres, and the
atoms located in the
plane $z$$=$$c/2$ are indicated by the light orange (grey) spheres.
The Mn-trimers, which transform to each other by the symmetry operation $\{ C^6_z | {\bf c}/2 \}$,
are shaded.}
\end{figure}

  The LMTO calculations have been performed for the AFM configuration
$\uparrow \downarrow \uparrow \downarrow \uparrow \downarrow$, where the arrows stand for the
directions of magnetic moments at the sites $1$-$6$ (see Fig.~\ref{fig.AtomicPositions} for the
notations of atomic positions).
The use of the AFM configuration is essential in order to open the band gap in LSDA
(about $0.7$ eV for YMnO$_3$, which is comparable with the experimental
optical gap of $1.3$ eV, reported in Ref.~\onlinecite{YMnO3_opt}).
Certain inconvenience of working with the
AFM $\uparrow \downarrow \uparrow \downarrow \uparrow \downarrow$ configuration is that
it artificially lowers the $P6_3cm$ symmetry: in this case, the local symmetry can by preserved
only around the sites $2$ and $5$, which will be selected as the reference points for the
analysis of interatomic magnetic interactions.

  In our LMTO calculations we decided to stick to the regular LSDA functional and
not to use any corrections for the on-site Coulomb interactions (LSDA$+$$U$). On the one hand, such corrections can
improve the description for interatomic magnetic interactions (similar to the model). On the other hand, the
use of the LSDA$+$$U$ functional is always conjugated with some additional uncertainties in the calculations, related to the
double-counting problem. Furthermore, the example of LaMnO$_3$ shows that LSDA is a reasonably good starting point for the
analysis of interatomic magnetic interactions.\cite{PRL96}
Nevertheless, when we compare the LMTO results with the
model calculations we discuss possible consequences of the Coulomb $U$ on the
magnetic interactions in the former case.

  Due to the hybridization with the oxygen states, which is treated explicitly in the LMTO calculations,
the value of spin magnetic moment at the Mn-sites is reduced till $3.5$ $\mu_{\rm B}$. Thus,
some deviation of the local magnetic moment from the ionic value ($4$ $\mu_{\rm B}$), which is typically
seen in the experiment,\cite{Park,Munoz} can be attributed to the covalent mixing. In model HF calculations,
similar effect can be described through the transformation
from the Wannier basis
to that of atomic orbitals.\cite{review2008}

\section{\label{Results} Results and Discussions}

\subsection{\label{MStructures} Optimization of Magnetic Structure}

  We start with the central result of our work and argue that the low-energy model (\ref{eqn.ManyBodyH}),
with the parameters derived from the first-principles electronic structure calculations,\cite{SM}
successfully reproduces the magnetic ground state of YMnO$_3$ and LuMnO$_3$.

\begin{figure}[h!]
\begin{center}
\includegraphics[width=15cm]{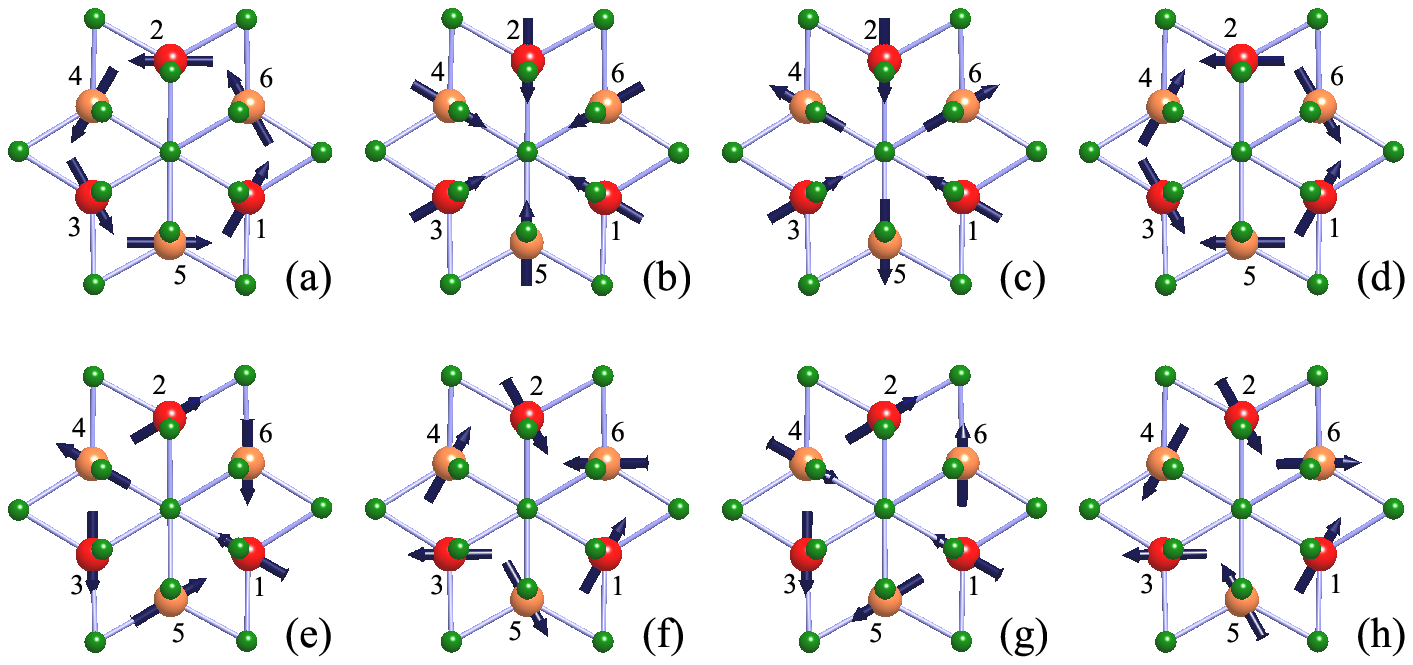}
\end{center}
\caption{\label{fig.MagneticStructures}
(Color online)
Magnetic structures obtained in the calculations (in the notations of Ref.~\protect\onlinecite{Munoz}):
$\Gamma_1$ (a), $\Gamma_2$ (b), $\Gamma_3$ (c), $\Gamma_4$ (d), $\Gamma_5$ with
${\bf e}_1 || [100]$ (e), $\Gamma_5$ with ${\bf e}_1 || [120]$ (f),
$\Gamma_6$ with ${\bf e}_1 || [100]$ (g), and $\Gamma_6$ with ${\bf e}_1 || [120]$ (h).
The oxygen atoms are indicated
by the small green (grey) spheres. The manganese atoms are indicated by the big spheres:
the ones located in the $z$$=$$0$ plane are shown by the red (dark) color, and the ones in the
$z$$=$$c/2$ plane -- by the light orange (grey) color.}
\end{figure}

  The main candidates for the magnetic ground state of YMnO$_3$ and LuMnO$_3$ are
shown in Fig.~\ref{fig.MagneticStructures}
(see also Refs.~\onlinecite{Munoz} and \onlinecite{Brown} for the notations).
The unidimensional representations $\Gamma_1$, $\Gamma_2$, $\Gamma_3$, and $\Gamma_4$
correspond to the magnetic space groups $P6_3cm$, $P6_3\underline{cm}$, $P\underline{6}_3c\underline{m}$,
and $P\underline{6}_3\underline{c}m$, respectively.
The directions of the spin magnetic moment, obtained in the HF
calculations for the low-energy model, are listed in Table~\ref{tab:Directions}.
\begin{table}[h!]
\caption{The angles $\alpha$ and $\beta$, representing the directions
${\bf e}_i = (\cos \alpha_i \cos \beta_i, \cos \alpha_i \sin \beta_i, \sin \alpha_i)$
of the spin magnetic moments in the plane $z$$=$$0$,
for different magnetic configurations (results of calculations using the experimental
crystal structure, measured at 10 K). The atomic positions are explained in Fig.~\protect\ref{fig.AtomicPositions}.
For the magnetic configurations $\Gamma_1$, $\Gamma_2$, and $\Gamma_6$,
the directions of the magnetic moments at the sites 4, 5, and 6 in the plane $z$$=$$c/2$ are obtained by the $180^\circ$
rotations around the $z$-axis of the vectors ${\bf e}_1$, ${\bf e}_2$, and ${\bf e}_3$.
For the magnetic configurations $\Gamma_3$, $\Gamma_4$, and $\Gamma_5$, these $180^\circ$ rotations should be combined with
the time inversion.}
\label{tab:Directions}
\begin{ruledtabular}
\begin{tabular}{ccc}
 configuration                        & YMnO$_3$                & LuMnO$_3$ \\
\hline
$\Gamma_1$ and $\Gamma_4$             &
$
\begin{array}{ll}
\alpha_1 = 0, & \beta_1 = 60^\circ  \\
\alpha_2 = 0, & \beta_2 = 180^\circ \\
\alpha_3 = 0, & \beta_3 = 300^\circ
\end{array}
$                                     &
$
\begin{array}{ll}
\alpha_1 = 0, & \beta_1 = 60^\circ  \\
\alpha_2 = 0, & \beta_2 = 180^\circ \\
\alpha_3 = 0, & \beta_3 = 300^\circ
\end{array}
$                                     \\
$\Gamma_2$ and $\Gamma_3$             &
$
\begin{array}{ll}
\alpha_1 = -$$0.2^\circ, & \beta_1 = 150^\circ  \\
\alpha_2 = -$$0.2^\circ, & \beta_2 = 270^\circ \\
\alpha_3 = -$$0.2^\circ, & \beta_3 = 30^\circ
\end{array}
$                                     &
$
\begin{array}{ll}
\alpha_1 = -$$0.2^\circ, & \beta_1 = 150^\circ  \\
\alpha_2 = -$$0.2^\circ, & \beta_2 = 270^\circ \\
\alpha_3 = -$$0.2^\circ, & \beta_3 = 30^\circ
\end{array}
$                                     \\
$\Gamma_5$ with ${\bf e}_1 || [100]$  &
$
\begin{array}{ll}
\alpha_1 = -$$9.6^\circ, & \beta_1 = 150^\circ  \\
\alpha_2 =    4.8^\circ, & \beta_2 = 30.5^\circ \\
\alpha_3 =    4.8^\circ, & \beta_3 = 269.5^\circ
\end{array}
$                                     &
$
\begin{array}{ll}
\alpha_1 = -$$8.8^\circ, & \beta_1 = 150^\circ  \\
\alpha_2 =    4.4^\circ, & \beta_2 = 30.3^\circ \\
\alpha_3 =    4.4^\circ, & \beta_3 = 269.8^\circ
\end{array}
$                                     \\
$\Gamma_5$ with ${\bf e}_1 || [120]$  &
$
\begin{array}{ll}
\alpha_1 =            0, & \beta_1 = 60^\circ  \\
\alpha_2 = -$$8.3^\circ, & \beta_2 = 299.5^\circ \\
\alpha_3 =    8.3^\circ, & \beta_3 = 180.5^\circ
\end{array}
$                                     &
$
\begin{array}{ll}
\alpha_1 =            0, & \beta_1 = 60^\circ  \\
\alpha_2 = -$$7.6^\circ, & \beta_2 = 299.8^\circ \\
\alpha_3 =    7.6^\circ, & \beta_3 = 180.3^\circ
\end{array}
$                                     \\
$\Gamma_6$ with ${\bf e}_1 || [100]$  &
$
\begin{array}{ll}
\alpha_1 = -$$13.4^\circ, & \beta_1 = 150^\circ  \\
\alpha_2 =     6.6^\circ, & \beta_2 = 30.8^\circ \\
\alpha_3 =     6.6^\circ, & \beta_3 = 269.2^\circ
\end{array}
$                                     &
$
\begin{array}{ll}
\alpha_1 = -$$23.3^\circ, & \beta_1 = 150^\circ  \\
\alpha_2 =    11.4^\circ, & \beta_2 = 30.1^\circ \\
\alpha_3 =    11.4^\circ, & \beta_3 = 268.0^\circ
\end{array}
$                                     \\
$\Gamma_6$ with ${\bf e}_1 || [120]$  &
$
\begin{array}{ll}
\alpha_1 =              0, & \beta_1 = 60^\circ   \\
\alpha_2 =  -$$11.5^\circ, & \beta_2 = 299.2^\circ \\
\alpha_3 =     11.5^\circ, & \beta_3 = 180.8^\circ
\end{array}
$                                     &
$
\begin{array}{ll}
\alpha_1 =              0, & \beta_1 = 60^\circ   \\
\alpha_2 =  -$$20.2^\circ, & \beta_2 = 297.8^\circ \\
\alpha_3 =     20.2^\circ, & \beta_3 = 182.2^\circ
\end{array}
$                                     \\
\end{tabular}
\end{ruledtabular}
\end{table}
In the $\Gamma_1$ and $\Gamma_4$ configurations, all magnetic moments lie in the $xy$-planes,
while in the $\Gamma_2$ and $\Gamma_3$ configurations, there is also a small canting of
magnetic moments along $z$. Moreover, the $\Gamma_2$ configuration allows for the
weak ferromagnetism along $z$, while in the $\Gamma_3$ configuration, the $z$-components of the
magnetic moments in the planes $z$$=$$0$ and $z$$=$$c/2$ cancel each other.
More generally, the configurations $\Gamma_1$ ($\Gamma_2$) and $\Gamma_4$ ($\Gamma_3$) differ by the magnetic alignment
in adjacent $xy$-planes: $\{ C^6_z | {\bf c}/2 \}$ acts as the normal symmetry operation in $\Gamma_1$ and $\Gamma_2$,
which transforms these states to themselves, while in
$\Gamma_3$ and $\Gamma_4$, $\{ C^6_z | {\bf c}/2 \}$ enters the magnetic symmetry group in the combination with the
time-inversion operation $\hat{T}$. It corresponds to the additional flip of magnetic moments
in the odd $xy$-planes of $\Gamma_3$ and $\Gamma_4$.
We have also considered other possible magnetic configurations with the symmetries $\Gamma_5$ and $\Gamma_6$,
as explained in Ref.~\onlinecite{Munoz}. However, as it will become clear from the discussion below,
they have higher energies.

The total energies of different magnetic configurations are summarized in Table~\ref{tab:TotalE}.
\begin{table}[h!]
\caption{Total energies of different magnetic configurations as obtained in the Hartree-Fock calculations
for the low-energy model. The energies are measured in meV per one formula unit, relative to the
most stable configuration. The magnetic
configurations are explained in Fig.~\protect\ref{fig.MagneticStructures}. The calculations for YMnO$_3$
and LuMnO$_3$ have been performed using the experimental crystal structure, measured at 10 K and 300 K
(as denoted in the notations).}
\label{tab:TotalE}
\begin{ruledtabular}
\begin{tabular}{ccccc}
 configuration                        & YMnO$_3$ (10 K) & YMnO$_3$ (300 K) & LuMnO$_3$ (10 K) & LuMnO$_3$ (300 K) \\
\hline
 $\Gamma_1$                           & $0.37$          & $0.20$           & $0.48$           & $0.23$            \\
 $\Gamma_2$                           & $0.16$          & $0.19$           & $0.61$           & $0.32$            \\
 $\Gamma_3$                           & $0$             & $0$              & $0.13$           & $0.10$            \\
 $\Gamma_4$                           & $0.21$          & $0.01$           & $0$              & $0$               \\
 $\Gamma_5$ with ${\bf e}_1 || [100]$ & $0.90$          & $0.76$           & $1.09$           & $1.06$            \\
 $\Gamma_5$ with ${\bf e}_1 || [120]$ & $0.90$          & $0.76$           & $1.09$           & $1.06$            \\
 $\Gamma_6$ with ${\bf e}_1 || [100]$ & $1.06$          & $0.94$           & $1.53$           & $1.27$            \\
 $\Gamma_6$ with ${\bf e}_1 || [120]$ & $1.06$          & $0.94$           & $1.53$           & $1.27$            \\
\end{tabular}
\end{ruledtabular}
\end{table}
Thus, the ground state of YMnO$_3$ is $\Gamma_3$
($P\underline{6}_3c\underline{m}$), in agreement with the experiment.\cite{Park,Fiebig2000}
In LuMnO$_3$, the ground state changes to $\Gamma_4$ ($P\underline{6}_3\underline{c}m$),
also in agreement with the experiment.\cite{Park,Fiebig2000}
However, all the states are located in a narrow energy range, which is expected for frustrated magnetic systems.
The lower-symmetry magnetic structure $P\underline{6}_3$,
which is typically regarded as another possible candidates for the magnetic ground state
of the hexagonal manganites,\cite{Park,Brown,Fiebig2000}
appears to be unstable and steadily converges
to either $P\underline{6}_3c\underline{m}$ (YMnO$_3$) or $P\underline{6}_3\underline{c}m$ (LuMnO$_3$).

  The band gap, obtained for both YMnO$_3$ and LuMnO$_3$, is about 2 eV,
which is larger than experimental 1.3 eV.\cite{YMnO3_opt}
Nevertheless, such a discrepancy is quite expectable for the level of the HF calculations.

\subsection{\label{MInteractions} Analysis of Magnetic Interactions}

  In this section, we clarify results of the HF calculations for the low-energy model and argue that
such a good agreement with the experimental data for the magnetic ground state is not surprising and can
be understood from the analysis of corresponding magnetic interactions,
which in turn depend on details of the lattice distortions in YMnO$_3$ and LuMnO$_3$.
Thus, we consider the spin model:
\begin{equation}
\hat{\cal H}_S = - \sum_{ \langle ij \rangle } J_{ij} {\bf e}_i {\bf e}_j +
\sum_{ \langle ij \rangle } {\bf d}_{ij} [{\bf e}_i \times {\bf e}_j] +
\sum_i {\bf e}_i
\hat{\tau}_i {\bf e}_i,
\label{eqn:SpinModel}
\end{equation}
which can be obtained by eliminating the electronic degrees of freedom form the more general
Hubbard model (\ref{eqn.ManyBodyH}), or directly from the
LMTO calculations.\cite{review2008,Liechtenstein,PRL96,DMI,MnCuN,PRB95}
In these notations, $\{ J_{ij} \}$ are the isotropic exchange interactions,
$\{ {\bf d}_{ij} \}$ are the antisymmetric DM interactions,
$\{ \hat{\tau}_i \}$ are the single-ion anisotropy tensors,
${\bf e}_i$ stands the \textit{direction} of the spin magnetic moment at the site $i$,
and the summation runs over all \textit{pairs} of atoms $\langle ij \rangle$.

   The parameters of isotropic magnetic interactions are listed in Table~\ref{tab:Isotropic}, and the
atomic positions are explained in Fig.~\ref{fig.AtomicPositions}.
\begin{table}[h!]
\caption{Parameters of isotropic exchange interactions (measured in meV),
calculated in the ferromagnetic states of YMnO$_3$ and LuMnO$_3$.
The atomic positions are explained in Fig.~\protect\ref{fig.AtomicPositions}.
Calculations have been performed using the experimental parameters of the crystal structure, measured at 10 K and 300 K
(as denoted in the notations).}
\label{tab:Isotropic}
\begin{ruledtabular}
\begin{tabular}{lrrrr}
 bond                        & YMnO$_3$ (10 K) & LuMnO$_3$ (10 K) & YMnO$_3$ (300 K) & LuMnO$_3$ (300 K) \\
\hline
 $2$-$1$                     & $-$$21.28$      & $-$$31.81$       & $-$$23.26$       & $-$$30.16$        \\
 $2$-$1'$                    & $-$$26.35$      & $-$$27.57$       & $-$$22.67$       & $-$$27.92$        \\
 $2$-$4$                     & $-$$0.12$       & $-$$0.20$        & $-$$0.13$        & $-$$0.20$         \\
 $2$-$5'$                    & $-$$0.19$       & $-$$0.11$        & $-$$0.08$        & $-$$0.10$         \\
 $2$-$4'$                    & $-$$0.24$       & $-$$0.31$        & $-$$0.21$        & $-$$0.24$         \\
 $2$-$5$                     & $-$$0.07$       & $-$$0.16$        & $-$$0.16$        & $-$$0.23$         \\
\end{tabular}
\end{ruledtabular}
\end{table}
All NN interactions in the plane $xy$ are AFM. This is
reasonable, because the ferromagnetic (FM) coupling in the hexagonal geometry can be stabilized only
by virtual hoppings onto unoccupied $3z^2$$-$$r^2$ orbital, which are relatively small (see Ref.~\onlinecite{SM}).
Moreover, the number of orbital paths, available for the virtual hoppings via this particular $3z^2$$-$$r^2$ orbital,
is also small. Nevertheless, from the orbital decomposition of $J_{ij}$ in our LMTO calculations, we can conclude that
such a FM contribution does exists and compensates about 30 \% of AFM contributions, involving all other
orbitals, except $3z^2$$-$$r^2$.

  The symmetry of the $P6_3cm$ lattice is such that there are two types of NN interactions.
The first type takes place in the triangles of atoms $1$-$2$-$3$ ($4$-$5$-$6$), which are
either expanded (the case of YMnO$_3$) or contracted (the case of LuMnO$_3$). The second type
takes place in the bonds $2$-$1'$, $2$-$3'$, $2$-$1''$, $2$-$3''$, which are all equivalent.
Then, due to the mirror reflection $x$$\rightarrow$$-$$x$, the NN bonds $2$-$4$ and $2$-$6$ between
adjacent $xy$-planes are also equivalent, and differ from the bond $2$-$5'$.
The same situation holds for the next-NN interactions between the planes: there are two equivalent
bonds $2$-$4'$ and $2$-$6'$, which differ from the bond $2$-$5$.
For the NN interactions, both in and between adjacent $xy$-planes, there is a clear correlation
between the bondlength and the strength of the exchange coupling. For example, in the low-temperature
structure of YMnO$_3$, the triangle of atoms $1$-$2$-$3$ ($4$-$5$-$6$) is \textit{expanded}
(for two inequivalent NN bonds $2$-$1$ and $2$-$1'$ in the $xy$-plane, the ratio of the bondlengths is
$l_{21'}/l_{21}$$=$$0.961$). Therefore, the AFM interaction $J_{21'}$ is stronger than
$J_{21}$.\cite{Park} The same tendency holds for the interplane interactions: for two inequivalent
NN bonds $2$-$5'$ and $2$-$4$ ($l_{25'}/l_{24}$$=$$0.991$), the AFM interaction
$J_{25'}$ is stronger than $J_{24}$. In LuMnO$_3$, where the triangle of atoms
$1$-$2$-$3$ ($4$-$5$-$6$) is \textit{compressed}, the situation is the opposite:
$l_{21'}/l_{21}$$=$$1.016$ and $l_{25'}/l_{24}$$=$$1.003$.
Therefore, the
exchange interactions in the bonds $2$-$1$ and $2$-$4$ are stronger than in the bonds
$2$-$1'$ and $2$-$5'$.

  The behavior of next-NN interactions between the planes obeys quite different rules.
Since the direct transfer integrals are small (see Ref.~\onlinecite{SM} for details), these
interactions are realized as the ``super-superexchange'' processes via intermediate sites in the pathes
$2$$\rightarrow$$6$$\rightarrow$$5$, $2$$\rightarrow$$1$$\rightarrow$$5$, etc., which always
include one compressed and one expanded bond. Therefore, the simple analysis in terms
of the bondlengths is no longer applicable. Instead,
we have found that for all considered compounds (and all considered
structures), the AFM interaction in the bond $2$-$5$ appears to be weaker than in the bonds
$2$-$4'$ (and in the equivalent to it bond $2$-$6'$).
Such a behavior has very important consequences: in a noncollinear structure,
it is more favorable energetically to form the FM coupling in the bond $2$-$5$
in order to maximize the AFM coupling in two other next-NN bonds $2$-$4'$ and $2$-$6'$.
Particularly, it explains why the magnetic ground state of YMnO$_3$ and LuMnO$_3$ should be
$\Gamma_3$, $\Gamma_4$, or $\Gamma_5$, which are characterized by the FM coupling in the bond $2$-$5$,
and not $\Gamma_1$, $\Gamma_2$ or $\Gamma_6$, where this coupling is AFM (see Fig.~\ref{fig.MagneticStructures}).
In LuMnO$_3$, this effect is additionally enhanced by the
NN interactions between the planes: since AFM interaction in the bond $2$-$5'$ is weaker than
in two equivalent bonds $2$-$4$ and $2$-$6$, it is more favorable energetically to form the FM coupling
between the sites $2$, $5$ and $5'$, where the latter two are connected by the translation.
However, in YMnO$_3$, the situation is
the opposite and there is a strong competition between NN and next-NN interactions between the planes.
Particularly, it explains a small energy difference between configurations $\Gamma_3$
and $\Gamma_2$.

  The reliability of the obtained parameters can be checked by calculating
$\theta_{\rm CW}$. In the case of classical Heisenberg model,
the latter is given by the formula $\theta_{\rm CW} \approx \sum_i J_{2i}/3 k_{\rm B}$, which yields
$-$$562$ and $-$$650$ K for the 10 K structure of YMnO$_3$ and LuMnO$_3$, respectively.
In the quantum case, these values should be additionally multiplied by $(1$$+$$1/S)$.
The structural changes have some effect mainly on YMnO$_3$, and, if one uses
parameters obtained for the 300 K structure,
$| \theta_{\rm CW} |$ decreases by 7\% (for comparison, similar change of $\theta_{\rm CW}$
for LuMnO$_3$ is about 1\%). In any case, the obtained values are in a good agreement with
experimental data.\cite{Park,LeeNature,Katsufuji}
The calculations of $T_{\rm N}$ are not straightforward: due to the
quasi-two-dimensional character of isotropic exchange interactions, $T_{\rm N}$ will be strongly
suppressed by thermal fluctuations, as one of the consequences of the Mermin-Wagner theorem.\cite{MerminWagner}
Of course, the molecular-field approximation will overestimate $T_{\rm N}$
(by factor 4, in comparison with the experiment).

  The LMTO calculations yield the following values of NN interactions in the plane $xy$ (in meV):
$(J_{21},J_{21'})$$=$ $(-$$12.8,-$$19.6)$, $(-$$18.0,-$$17.0)$, $(-$$15.8,-$$16.6)$, and $(-$$17.0,-$$17.0)$
for YMnO$_3$ (10 K), LuMnO$_3$ (10 K), YMnO$_3$ (300 K), and LuMnO$_3$ (300 K), respectively.
Thus, all the interactions are weaker than in the model analysis. Nevertheless, this seems reasonable.
First, the NN interactions are generally weaker in the AFM $\uparrow \downarrow \uparrow \downarrow \uparrow \downarrow$
configuration. This effect was also found in the model calculations, as will become clear below. Second, the
ratio of AFM to FM contributions to the exchange coupling in manganites scales with the value of $U$ as
$(U$$-$$J_{\rm H})/(U$$+$$3J_{\rm H}) \approx 1$$-4$$J_{\rm H}/U$.\cite{KugelKhomskii}
Thus, larger $U$, which was used in the model (but not in the LMTO calculations), will shift this balance
towards the AFM coupling. Similar tendency was found for inter-layer interactions: although LSDA, supplementing
the LMTO calculations, somewhat overestimates FM contributions to the exchange interactions,
the modulation of these interactions, caused by the lattice distortion, again favors the
formation of magnetic configurations
$\Gamma_3$ or $\Gamma_4$.
For example, in YMnO$_3$ (10 K), the LMTO calculations yield: $J_{24}$$= 0.20$ meV,
$J_{25'}$$= 0.04$ meV, $J_{24'}$$=-$$0.14$ meV, and $J_{25}$$= 0.08$ meV.
Therefore, these calculations confirm that the experimental coupling between hexagonal layers
is stabilized by the next-NN interactions $J_{25} > J_{24'}$. The NN interactions act
in the opposite direction: $J_{25'} < J_{24}$. However, their effect is smaller.

  Let us discuss the behavior of the single-ion anisotropy tensor.
Due to the mirror reflection $x$$\rightarrow$$-$$x$, the tensor $\hat{\tau}_2$ at the
site 2 (see Fig.~\ref{fig.AtomicPositions}) has the following form:
$$
\hat{\tau}_2 =
\left(
\begin{array}{ccc}
\tau^{xx} & 0         &  0        \\
0         & \tau^{yy} & \tau^{yz} \\
0         & \tau^{zy} & \tau^{zz} \\
\end{array}
\right),
$$
where $\tau^{zy} = \tau^{yz}$ and $\tau^{xx}$$+$$\tau^{yy}$$+$$\tau^{zz} = 0$. Thus, the magnetic
moments can either lie along the $x$-axis or be perpendicular to it. In the latter case (and if
$\tau^{yz}$$\ne$$0$) they from a canted magnetic structure. The anisotropy tensors at other Mn-sites
can be generated by applying the symmetry operations of the space group $P6_3cm$. The matrix elements
of $\hat{\tau}_2$ can be evaluated in the second order of perturbation theory expansion with respect to the
SO interactions.\cite{PRB95} Then, near the FM state, we obtain the following sets of
independent parameters (in meV):
$(\tau^{yy},\tau^{yz},\tau^{zz})$$=$ $(-$$0.34,-$$0.12,0.58)$,
$(-$$0.29, -$$0.11,0.58)$, $(-$$0.25,-$$0.12,0.57)$, and
$(-$$0.26,-$$0.12,0.57)$ for YMnO$_3$ (10 K), YMnO$_3$ (300 K),
LuMnO$_3$ (10 K), and LuMnO$_3$ (300 K), respectively.
Since $\tau^{zz}$$>$$\tau^{yy}$, all structures
with large $z$-components of
the magnetic moments are energetically unfavorable.
Then, by diagonalizing $\hat{\tau}_2$, one can find that
the lowest-energy configuration in LuMnO$_3$ is the one where the
magnetic moment at the site 2 is parallel to the $x$-axis. The next, canted magnetic configuration,
is higher in energy by about $0.05$ meV
(for the 10 K structure).
This situation is reversed in YMnO$_3$, where the lowest energy corresponds
to the canted magnetic configuration. The angle $\alpha$, formed by
the magnetic moment and the $y$-axis,
is about $7^\circ$. In the next configuration,
which is higher in energy by about $0.10$ meV (for the 10 K structure),
the magnetic moment is parallel to the $x$-axis.
This energy difference is reduced till $0.01$ meV for the 300 K structure.
The same behavior was found in the LMTO calculations: for YMnO$_3$, the lowest energy
corresponds to the canted magnetic configuration (the canting from the $y$-axis is
about $6^\circ$). The next configuration, where the magnetic moment is parallel
to the $x$-axis, is higher in energy by $0.09$ eV for the 10 K structure, and this
energy difference further decreases for the 300 K structure.

  Thus, the change of the ground state from $\Gamma_3$ to $\Gamma_4$ in the direction from
YMnO$_3$ to LuMnO$_3$ is related to the behavior of the single-ion anisotropy,
which in turns correlates with the distortion of the $1$-$2$-$3$ triangles
(expansion and contraction, respectively). Moreover, due to the $180^\circ$ rotation
around the $z$-axis, which is required in order to transform the site $2$ to the site $5$
(see Fig.~\ref{fig.AtomicPositions}), the matrix element $\tau^{yz}$ will change sign.
Therefore, the canting of spins in the planes $z$$=$$0$ and $z$$=$$c/2$
of the $\Gamma_3$ structure will act in the opposite directions, and
the magnetic moments along the $z$-axis will cancel each other.

  The single-ion anisotropy will tend to align $z$-components of the magnetic moments
ferromagnetically in each of the $xy$-plane. However, this effect will compete with
the NN AFM interactions $J_{21}$ and $J_{21'}$. The corresponding analytical expression for the
spin canting can be obtained by minimizing the energies of single-ion anisotropy and
isotropic exchange interactions: by assuming that all neighboring spins in the $xy$-plane form the
$120^\circ$-structure (as in the case of the $\Gamma_2$ and $\Gamma_3$ configurations),
one can find that
\begin{equation}
\tan 2 \alpha = - \frac{ 2 \tau^{yz} } { \tau^{yy}-\tau^{zz} + 3 J_{21} + 6 J_{21'} },
\label{eqn:canting}
\end{equation}
where the minus-sign corresponds to the situation, which is realized in our HF calculations and
where ${\bf e}_2$ is antiparallel to the $y$-axis (see Fig.~\ref{fig.MagneticStructures}).
Then, for the $\Gamma_3$ configuration of YMnO$_3$ (10 K), the canting angle $\alpha$ can be estimated
(using both model and LMTO parameters of magnetic interactions) as
$\alpha \approx - \tau^{yz}/(3J_{21} + 6J_{21'}) = -$$0.03^\circ$, which is about 7 times smaller
than the values obtained in
self-consistent HF calculations (Table~\ref{tab:Directions}). Nevertheless,
there is an additional contribution to the spin canting, caused by
the DM interactions.

  Parameters of DM interactions between NN sites in the $xy$-plane are listed in Table~\ref{tab:DM}.
\begin{table}[h!]
\caption{Parameters of Dzyaloshinskii-Moriya interactions (measured in meV),
calculated in the ferromagnetic states of YMnO$_3$ and LuMnO$_3$.
The atomic positions are explained in Fig.~\protect\ref{fig.AtomicPositions}.
Calculations have been performed using the experimental parameters of the crystal structure, measured at 10 K and 300 K
(as denoted in the notations).}
\label{tab:DM}
\begin{ruledtabular}
\begin{tabular}{lcccc}
 bond                        & YMnO$_3$ (10 K)       & LuMnO$_3$ (10 K)      & YMnO$_3$ (300 K)      & LuMnO$_3$ (300 K) \\
\hline
 $2$-$1$                     & $(0.01,0.01,0.20)$    & $(0.04,0.02,0.25)$    & $(0.04,0.02,0.17)$    &  $(0.07,0.04,0.25)$        \\
 $2$-$1'$                    & $(0.03,-$$0.02,0.21)$  & $(0.03,-$$0.02,0.26)$ & $(0.03,-$$0.01,0.18)$ & $(0.02,-$$0.01,0.26)$     \\
 $2$-$1''$                   & $(0,0.04,0.21)$       & $(-$$0.01,0.04,0.26)$ & $(0,0.03,0.18)$       & $(0,0.02,0.26)$        \\
\end{tabular}
\end{ruledtabular}
\end{table}
They were obtained by considering mixed perturbation theory expansion with respect to the SO interaction
and infinitesimal rotations of spin magnetic moments.\cite{PRL96}
In principle, the parameters ${\bf d}_{21'}$ and ${\bf d}_{21''}$ are not independent and can be
transformed to each other using symmetry operations of the space group $P6_3cm$. However, it is more
convenient to consider their contributions independently.
Due to the mirror reflection $x$$\rightarrow$$-$$x$, the elements of two \textit{axial} vectors
${\bf d}_{23}$ and ${\bf d}_{21}$ (see Fig.~\ref{fig.AtomicPositions}) obey the following rules:
$d^x_{23} = d^x_{21}$, $d^y_{23} =-$$d^y_{21}$, and $d^z_{23} =-$$d^z_{21}$
(similar situation holds for other NN interactions). Thus, they will produce a finite canting
at the site $2$ only if the directions of two other magnetic moments ${\bf e}_2$ and ${\bf e}_3$
would have an AFM component along $x$ and a FM component along $y$, i.e.:
$e^x_3 = -$$e^x_1$ and $e^y_3 = e^y_1$. Such a situation
is realized in the magnetic configurations $\Gamma_2$ and $\Gamma_3$
(but not in $\Gamma_1$ and $\Gamma_4$). Then, the magnetic moment at the site 2
will experience the additional rotational force from the sites $1$, $1'$, and $1''$:
${\bf f}_{1 \to 2} = [ {\bf d}_{21} \times {\bf e}_1 ]$$+$$[ {\bf d}_{21'} \times {\bf e}_1 ]$$+$$
[ {\bf d}_{21''} \times {\bf e}_1 ]$. For the magnetic configurations $\Gamma_2$ and $\Gamma_3$,
the sites of the type `$3$' will create the same rotational force:
${\bf f}_{3 \to 2} = {\bf f}_{1 \to 2}$. However, for the $\Gamma_1$ and $\Gamma_4$ configurations,
it holds ${\bf f}_{3 \to 2} = -$${\bf f}_{1 \to 2}$. Therefore, these two contribution will
cancel each other and there will be no canting of spins.

  These rotational forces should be
incorporated in the expression (\ref{eqn:canting}) for the spin canting, which yields
$\alpha \approx -(\tau^{yz} + f_{1 \to 2}^z)/(3J_{21} + 6J_{21'}) = -$$0.04^\circ$.
This canting is still smaller than
$\alpha \sim -$$0.21^\circ$, obtained in the HF calculations for the $\Gamma_3$
configuration (Table~\ref{tab:Directions}).
Nevertheless, it should be noted that all the parameters of the spin Hamiltonian (\ref{eqn:SpinModel})
were evaluated using perturbation theory expansion near the collinear FM state, which is very
far from the ground-state configuration $\Gamma_3$. Thus, it is difficult to expect that the
perturbation theory, although is very useful for the semi-quantitative analysis, should be
able to reproduce all details of the solutions of the electronic model (\ref{eqn.ManyBodyH}). In fact, some parameters
of the spin Hamiltonian (\ref{eqn:SpinModel}) appear to be sensitive to the state, in which they are calculated.
For example, we have also considered the collinear AFM configuration
$\uparrow \downarrow \uparrow \downarrow \uparrow \downarrow$, where the arrows stand for the
directions of magnetic moments at the sites $1$-$6$. In this case, the DM interactions involving the
site $2$, which is AFM coupled with all NN spins in the $xy$-plane, become (in meV):
${\bf d}_{21} = (0.01,0.03,0.01)$, ${\bf d}_{21'} = (0.04,0,0)$, and
${\bf d}_{21''} = (-$$0.01,0.05,0.01)$.
Then, corresponding rotational force $f_{1 \to 2}^z$ will be about 2 times larger than in the FM state.
Meanwhile, the parameters of isotropic exchange interactions
$J_{21}$ and $J_{21'}$ decrease by about 15\%. These factors will additionally increase $\alpha$.

  Furthermore, the
HF potential for the low-energy model (\ref{eqn.ManyBodyH}) is orbitally dependent. In this cases, the
local force theorem is no longer valid.\cite{Liechtenstein} Therefore, the total energy change due to the SO interaction
can be replaced
only approximately by the change of the single-particle energies of the HF method.
For the single-ion anisotropy, the situation was discussed in Appendix B of Ref.~\onlinecite{PRB95}.
Presumably, this is the main reason, explaining the quantitative difference between the
results of the electronic and spin models.
Thus, these are typical uncertainties, supplementing the
construction and analysis of the spin model (\ref{eqn:SpinModel}).

  Nevertheless, the local force theorem is valid within LSDA. Therefore, it is interesting to
estimate the spin canting in the LMTO calculations, which are based on the LSDA functional.
In this case, all DM interactions become larger. For example, for YMnO$_3$ (10 K) we have obtained
the following parameters (in meV): ${\bf d}_{21} = (-$$0.01,0.14,-$$0.20)$,
${\bf d}_{21'} = (-$$0.16,0.04,-$$0.12)$, and ${\bf d}_{21''} = (0.06,0.18,-$$0.26)$.
Then, by combining them with corresponding parameters of the single-ion anisotropy
$\tau^{yz} = -$$0.078$ meV and isotropic exchange interactions $J_{21}$ and $J_{21'}$,
which are listed above,
we obtain the canting angle $\alpha = -$$0.12^\circ$.
Thus, it is interesting that LSDA, despite its limitation, provides the best starting point
for the analysis of the spin canting via the perturbation-theory expansion for the
spin-orbit interaction, due to validity of the local force theorem.
Similar situation was found in the orthorhombic LaMnO$_3$.\cite{PRL96}

  Thus, although derivation of parameters
of spin model (\ref{eqn:SpinModel}) may differ in details, depending on the
form of the electronic Hamiltonian, which is used as the starting point, as well as
some additional approximations, underlying definitions of the model parameters,
this analysis provides a clear microscopic basis for understanding the main difference
between YMnO$_3$ and LuMnO$_3$: why the former tends to form the canted noncollinear
magnetic structure $\Gamma_3$, while the latter forms the planar structure $\Gamma_4$.

\subsection{\label{Polarization} Magnetic Contribution to Ferroelectric Polarization}

  Finally, we would like to comment on the behavior of electronic polarization ${\bf P} || {\bf c}$.
It was calculated within the Berry-phase formalism,\cite{BerryPhase} which was adopted for the model calculations.\cite{PRB11}
Of course, the ferroelectric activity in YMnO$_3$ and LuMnO$_3$ is primarily caused by structural effects.
For example, in YMnO$_3$, the ferroelectric transition occurs at about $T_{\rm C} = 880$ K,\cite{ChoiNM}
which is much higher than $T_{\rm N} = 75$ K.\cite{LeeNature} This fact was also confirmed by
first-principles calculations.\cite{vanAkenNM} Another appealing evidence is that the ferroelectric
domains in YMnO$_3$ always coincide with the structural ones.\cite{ChoiNM}
Nevertheless, beside this
structural deformation, we have found that there is a substantial magnetic contribution to ${\bf P} || {\bf c}$.
More specifically, all magnetic configurations can be divided in two group. The first one includes
$\Gamma_1$, $\Gamma_2$, and $\Gamma_6$, where the magnetic moments in the planes
$z$$=$$0$ and $z$$=$$c/2$
can be transformed to each other by the simple rotations.
The second groups includes $\Gamma_3$, $\Gamma_4$, and $\Gamma_5$, where these rotations should be
additionally combined with the time inversion. According to our finding,
the states in each group are characterized by nearly
equal values of ${\bf P} || {\bf c}$. However,
the transformation of the magnetic state from one group to another would cause a finite jump
of electronic polarization. Thus, in principle, the value of the ferroelectric polarization can be controlled by changing
the magnetic state (and vice versa). In this sense, more promising candidate is YMnO$_3$, where
the ground state ($\Gamma_3$) and the first excited state ($\Gamma_2$) belong to different groups.
The energy difference $\Delta E$ between these two configurations is about $0.16$ meV (see Table~\ref{tab:TotalE}).
Then, the change of the ferroelectric polarization, associated with the change of the magnetic state
$\Gamma_3$$\rightarrow$$\Gamma_2$, can be estimated as
$\Delta {\bf P} || {\bf c}$$= -$$120$ $\mu$C/m$^2$.
The practical realization of such a switching phenomenon
would be probably interesting, although it is not immediately clear, which external interaction
could switch the magnetic state.
Formally speaking, the magnetic configuration $\Gamma_2$ could be stabilized
by the external electric field ${\bf E} || {\bf c}$, which couples to $\Delta {\bf P}$
and results in the additional energy gain $-$$\Delta {\bf P} {\bf E}$. Alternatively,
one could exploit the fact that $\Gamma_2$ allows for a weak ferromagnetism along $z$
(while $\Gamma_3$ does not) and, therefore, could be also stabilized by
interaction with the external magnetic field, $-$$\Delta {\bf M} {\bf B}$, which couples
to the net magnetic moment $\Delta {\bf M}$ ($\sim -$$0.01 \mu_{\rm B}$ per Mn-site).
However, in order to overcome the total energy difference $\Delta E$,
this would require unrealistically large values of ${\bf E}$ and ${\bf B}$, which cannot be realized in practice.
Therefore, one should explore alternative possibilities.
For example,
from the viewpoint of microscopic interactions, one could use
the competition of the NN and next-NN interactions
between adjacent $xy$-planes, which in the case of YMnO$_3$ act in the \textit{opposite} direction
(see discussions above). The $\Gamma_3$ configuration is stabilized by the next-NN interactions.
However, if one could find such macroscopic conditions, which would shift this balance towards
NN interactions, one could switch the magnetic structure $\Gamma_3$$\rightarrow$$\Gamma_2$ and,
therefore, the ferroelectric polarization. Another possibility is, of course, to exploit
magnetism of the rare-earth ions, which can act similar to ${\bf B}$, but produces much stronger
effect on the Mn-sublattice. Such a magnetic phase control was indeed realized experimentally
in the series of hexagonal manganites with the magnetic rare-earth sublattices.\cite{Fiebig2004,Fiebig2003}

\section{\label{Conclusions} Summary}

  Using results of first-principles electronic structure calculations,
we have established the low-energy model, which is able to capture basic magnetic properties of hexagonal manganites.
This Hubbard-type model describes the behavior of the Mn $3d$ bands, being subjected to the lattice deformation and
on-site
electron-electron interactions. All parameters of such model,
derived from the first principles calculations for two characteristic manganites
YMnO$_3$ and LuMnO$_3$, are summarized in Ref.~\onlinecite{SM}.

  Then, the model was solved in the HF approximation, by considering
all possible noncollinear magnetic structures with different symmetries.
Since the
magnetic frustration in the hexagonal $P6_3cm$ lattice is lifted by the relativistic SO interaction,
the HF approximation provides a good starting point for the analysis of the magnetic properties
of these compounds and
successfully reproduce
the experimental change of the magnetic ground state
from $P\underline{6}_3c\underline{m}$ to $P\underline{6}_3\underline{c}m$ in the direction
from YMnO$_3$ to LuMnO$_3$, which was observed in the neutron diffraction and nonlinear optical studies.

  In order to clarify the microscopic origin of such a change, we have further transformed the
electronic model into the spin one and discussed the same trend in terms of
differences in the behavior of magnetic interactions in these systems.
We have found that the main reason why YMnO$_3$ and LuMnO$_3$
tend to form different magnetic structure
is related to the behavior of the single-ion anisotropy,
which couples to the trimerization distortion in the hexagonal plane
and reflects different directions of this trimerization in YMnO$_3$ and LuMnO$_3$ (expansion and construction
of the Mn-trimers, respectively).
On the other hand, the interplane coupling in both compounds is controlled by the next-NN interactions,
which is less sensitive to the direction of trimerization. The spin canting in the $P\underline{6}_3c\underline{m}$
structure of YMnO$_3$ is a combined effect of both single-ion anisotropy and Dzyaloshinskii-Moriya interactions,
which act in the same direction.
As the trimerization distortion decreases with the
temperature, all anisotropic interactions also decrease, thus reviving
the magnetic frustration and the degeneracy of the magnetic state.

  Finally, using the Berry-phase formalism, we have estimated the magnetic contribution
to the ferroelectric polarization and discussed how
it can be controlled by changing the magnetic structure of YMnO$_3$.

  \textit{Acknowledgements}.
The work of MVV and VVM is supported by the grant program of President of Russian Federation
 MK-406.2011.2, the scientific program ``Development of scientific potential of Universities''.


\begin{thebibliography}{99}

\bibitem{Mostovoy}
S.-W. Cheong and M. Mostovoy,
Nature Materials \textbf{6}, 13 (2007).

\bibitem{Fiebig2004}
Th. Lottermoser, Th. Lonkai, U. Amann, D. Hohlwein, J\"{o}rg, and M. Fiebig,
Nature \textbf{430}, 541 (2004).

\bibitem{Huang}
Z.~J. Huang, Y. Cao, Y.~Y. Sun, Y.~Y. Xue, C.~W. Chu,
Phys. Rev. B \textbf{56}, 2623 (1997).

\bibitem{Katsufuji}
T. Katsufuji, S. Mori, M. Masaki, Y. Moritomo, N. Yamamoto, and H. Takagi,
Phys. Rev. B \textbf{64}, 104419 (2001).

\bibitem{ChoiNM}
T. Choi, Y. Horibe, H.~T. Yi, Y.~J. Choi, W. Wu, and S.-W. Cheong,
Nature Materials \textbf{9}, 253 (2010).

\bibitem{Fiebig2002}
M. Fiebig, Th. Lottermoser, D. Fr\"olich, A.~V. Goltsev, and R.~V. Pisarev,
Nature \textbf{419}, 818 (2002).

\bibitem{LeeNature}
S. Lee, A. Pirogov, M. Kang, K.-H. Jang, M. Yonemura, T. Kamiyama, S.-W. Cheong,
F. Gozzo, N. Shin, H. Kimura, Y. Noda, and J.-G. Park,
Nature \textbf{451}, 805 (2008).

\bibitem{PRB11}
I.~V. Solovyev and Z.~V. Pchelkina,
Phys. Rev. B \textbf{82}, 094425 (2010);
I.~V. Solovyev,
\textit{ibid.} \textbf{83}, 054404 (2011).
Note that a prefactor was missing in the previous model calculations of ${\bf P}$,
and all values of the electric polarization, reported in this paper, should be additionally divided roughly by
$2.5$. This partly resolves the problem of disagreement with the experimental data.
The details will be discussed in a separate publication.

\bibitem{Perring}
J. Park, J.-G. Park, G. S. Jeon, H.-Y. Choi, C. Lee, W. Jo, R. Bewley, K.~A. McEwen, and T.~G. Perring,
Phys. Rev. B \textbf{68}, 104426 (2003).

\bibitem{Sato}
T.~J. Sato, S.-H. Lee, T. Katsufuji, M. Masaki, S. Park, J.~R.~D. Copley, and H. Takagi,
Phys. Rev. B \textbf{68}, 014432 (2003).

\bibitem{Park}
J. Park, S. Lee, M. Kang, K.-H. Jang, C. Lee, S.~V. Streltsov, V.~V. Mazurenko, M.~V. Valentyuk,
J.~E. Medvedeva, T. Kamiyama, and J.-G. Park,
Phys. Rev. B \textbf{82}, 054428 (2010).

\bibitem{Okuyama}
D. Okuyama, S. Ishiwata, Y. Takahashi, K. Yamauchi, S. Picozzi, K. Sugimoto, H. Sakai,
M. Takata, R. Shimano, Y. Taguchi, T. Arima, and Y. Tokura,
Phys. Rev. B \textbf{84}, 054440 (2011).

\bibitem{Munoz}
A. Mu\~noz, J.~A. Alonso, M.~J. Mart\'inez-Lope, M.~T. Cas\'ais, J.~L. Mart\'inez, and
M.~T. Fern\'andez-D\'iaz,
Phys. Rev. B \textbf{62}, 9498 (2000).

\bibitem{Brown}
P.~J. Brown and T. Chatterji,
J. Phys.: Condens. Matter \textbf{18}, 10085 (2006).

\bibitem{Fiebig2000}
M. Fiebig, D. Fr\"olich, K. Kohn, St. Leute, Th. Lottermoser, V.~V. Pavlov, and R.~V. Pisarev,
Phys. Rev. Lett. \textbf{84}, 5620 (2000).

\bibitem{review2008}
I.~V. Solovyev,
J. Phys.: Condens. Matter \textbf{20}, 293201 (2008).

\bibitem{SM}
Supplemental materials
[parameters of the crystal field, transfer integrals, and matrices
of Coulomb interactions].

\bibitem{LMTO}
O.~K. Andersen, Z. Pawlowska, and O. Jepsen,
Phys. Rev. B {\bf 34}, 5253 (1986).

\bibitem{Liechtenstein}
A.~I. Liechtenstein, M.~I. Katsnelson, V.~P. Antropov, and V.~A. Gubanov,
J. Magn. Magn. Matter. \textbf{67} 65 (1987).

\bibitem{PRL96}
I. Solovyev, N. Hamada, and K. Terakura,
Phys. Rev. Lett. \textbf{76}, 4825 (1996).

\bibitem{DMI} V.~V. Mazurenko and V.~I. Anisimov,
Phys. Rev. B \textbf{71}, 184434 (2005).
Note that this work employed the same strategy
for derivation of parameters of DM interactions
as in Ref.~\onlinecite{PRL96}, but different choice of phases
in the spin-rotation matrix. Namely,
the rotation of spin from
${\bf e}^0 = (0, 0, 1)$ to
${\bf e} = (\cos \varphi \sin \theta, \sin \varphi \sin \theta, \cos \theta)$
was described by the following sets
of the Euler angles: $(\alpha_I, \beta_I, \gamma_I) = (\varphi,\theta,-$$\varphi)$,
in this work, and $(\alpha_{II}, \beta_{II}, \gamma_{II}) = (\varphi,\theta,0)$,
in Ref.~\onlinecite{PRL96}. The second choice provides more compact expression
for DM interactions and allows to get rid of the on-site contribution to the
rotational force. Of course, the total force, created by all spins,
does not depend on the phase choice.

\bibitem{MnCuN}
A.~N. Rudenko, V.~V. Mazurenko, V.~I. Anisimov, A.~I. Lichtenstein,
Phys. Rev. B \textbf{79}, 144418 (2009).

\bibitem{PRB95}
I.~V. Solovyev, P.~H. Dederichs, and I. Mertig,
Phys. Rev. B \textbf{52}, 13419 (1995).

\bibitem{YMnO3_opt}
A.~M. Kalashnikova and R.~V. Pisarev,
JETP Letters {\bf 78}, 143 (2003).



\bibitem{MerminWagner}
N.~D. Mermin and H. Wagner,
Phys. Rev. Lett. \textbf{17}, 1133 (1966);
\textit{ibid.} \textbf{17}, 1307(E) (1966).

\bibitem{KugelKhomskii}
K.~I. Kugel and D.~I. Khomskii,
Sov. Phys. Usp. \textbf{25}, 231 (1982).

\bibitem{BerryPhase}
D. Vanderbilt and R.~D. King-Smith, Phys. Rev. B \textbf{48}, 4442 (1993);
R. Resta, J. Phys.: Condens. Matter \textbf{22}, 123201 (2010).


\bibitem{vanAkenNM}
B.~B. van Aken, T.~T.~M. Palstra, A. Filippetti, and N.~A. Spaldin,
Nature Materials \textbf{3}, 164 (2004).


\bibitem{Fiebig2003}
M. Fiebig, Th. Lottermoser, and R.~V. Pisarev,
J. Appl. Phys. \textbf{93}, 8194 (2003).



\end{thebibliography}
\end{document}